# Temperature mapping of stacked silicon dies from x-ray diffraction intensities

*Darshan Chalise [1,2], Peter Kenesei[3], Sarvjit D. Shastri[3], David G. Cahill[1,2,4]*

[1]*Department of Physics, University of Illinois at Urbana-Champaign, Urbana, IL, 61801, USA*

[2]*Materials Research Laboratory, University of Illinois at Urbana-Champaign, Urbana, IL, 61801, USA*

[3]*X-ray Science Division, Argonne National Laboratory, Lemont, Illinois 60439, USA*

[4]*Materials Science and Engineering, University of Illinois at Urbana-Champaign, Urbana, IL, 61801, USA*

*Corresponding Author – d-cahill@illinois.edu

## Abstract

Increasing power densities in integrated circuits has led to an increased prevalence of thermal hotspots in integrated circuits. Tracking these thermal hotspots is imperative to prevent circuit failures. In 3D integrated circuits, conventional surface techniques like infrared thermometry are unable to measure 3D temperature distribution and optical and magnetic resonance techniques are difficult to apply due to the presence of metals and large current densities. X-rays offer high penetration depth and can be used to probe 3D structures. We report a method utilizing the temperature dependence of x-rays diffraction intensity via the Debye-Waller factor to simultaneously map the temperature of an individual silicon die that is a part of a stack of dies. Utilizing beamline 1-ID-E at the Advanced Photon Source (Argonne), we demonstrate for each individual silicon die, a temperature resolution of 3 K, a spatial resolution of 100 μm x 400 μm and a temporal resolution of 20 s. Utilizing a sufficiently high intensity laboratory source, e.g., from a liquid anode source, this method can be scaled down to laboratories for non-invasive temperature mapping of 3D integrated circuits.

## I. INTRODUCTION

The increase in power density of processors has led to an increase in the occurrence of thermal hot spots within the processors [1,2]. The location of the thermal spots change with time, and therefore tracking them is difficult. Tracking of the hot spots, however, is important as most failure mechanisms in the processors have strong temperature dependence [1].

Traditionally, the processor dies are embedded with thermal sensors to measure the temperature during operation. However, these sensors provide only single point temperature measurements and are thus limited in spatial resolution. To map the temperature across the entire processor, non-contact thermometry is required [1].

Infrared thermometry can be used to measure the temperature distribution of the surface of a single die [2]. However, with 3D integrated circuits, infrared thermometry is unable to measure a 3D temperature field, especially because the scattering of infrared radiation from interfaces makes interpretation of observed radiation that is emitted from the surface of the sample extremely challenging. Furthermore, silicon is only transparent in the infrared range of wavelength 1.2 µm to 15 µm [3], and therefore, detecting infrared radiation from inner dies to reconstruct a temperature field is impractical. Optical techniques like thermoreflectance imaging [4] or optical coherence tomography (OCT) [5] would also be impractical due to strong absorption of visible or near-infrared light by metal-silicon composite even with a small amount of metal [6].

High energy x-rays have great penetration depth for most materials [7], and therefore offer a possibility for non-invasively probing structures like stacked silicon dies. For example, a 30 keV x-ray beam has a penetration depth of 100 µm in copper and 3 mm in silicon [7], and therefore, it could be used to probe ~10 layers (normal incidence) of a composite structure with 10 µm copper and 100 µm silicon. At 70 keV (~ energy used in the current experiment) the penetration depth increases to 1 mm in copper and 6 cm in silicon [7], and significantly thicker samples can be probed.

In general, several features of x-ray diffraction depend on temperature. X-ray peaks shift due to the changes in lattice parameter, the structure factor changes due to thermally-excited lattice vibration (Debye-Waller factor), and peaks can broaden due to inhomogeneous strain induced by differences in thermal expansion of adjoining materials [8]. Therefore, measurement of x-ray intensity provides a means of non-invasive thermometry of stacked structures like 3D integrated circuits.

Sensitivity analysis can be used to determine which x-ray method has the highest temperature dependence. The sensitivity coefficient, given by $S_T = \frac{\partial \ln S}{\partial \ln T}$, measures the percentage change in signal, S, with respect to 1% change in temperature, T [9]. In measuring the change in lattice parameter due to temperature, the lattice parameter changes by about $6 \times 10^{-6}$ radians for 1% (3 K) change in temperature. Therefore, measurements with a temperature resolution of 1 K would require the resolution of a Bragg center by ~2 µrad. The sensitivity coefficient of the measurement of thermally induced strain is similar because of the small coefficient of thermal expansion of materials, which are ~$2 - 3 \times 10^{-6}$ /K [10].

The measurement of Debye-Waller factor via the reduction of diffraction intensity provides a much better sensitivity in temperature measurement [11] (As discussed below, we predict and observe $S_T$ ~ 2.5 for the measurement of (16 0 0) diffraction peak for silicon). Therefore, this experiment proposes a method of measuring the relative change in Debye-Waller factor to measure the temperature of different layers of stacked silicon dies. The experiment utilizes rastering of an x-ray beam across the sample to produce a map of diffraction intensities, and thereby of temperature. Simultaneous mapping of different dies in a stack is possible due to the spatial separation of beams that are diffracted from different dies in the stack.

## II. THEORY

Thermal vibrations modify the structure factor (scattering amplitude) of the unit cell, which can be described by the introduction of a temperature factor, $e^{-M}$ [12], i.e.

$$|F_{mod}| = |F|e^{-M}$$

Where, $= \frac{2\pi(s)^2 \Delta X^2}{3}$; $\Delta X^2$ is the average quadratic displacement of an atom from its mean position, and $s$ is the scattering vector. $s = \frac{2\sin\theta}{\lambda}$ in the case of Bragg reflection, with θ as the Bragg angle and λ as the incident wavelength.

$M$ can be calculated approximately within the Debye model:

$$M = \frac{6h^2 T \left[\psi(x) + \frac{x}{4}\right] \left(\frac{\sin\theta}{\lambda}\right)^2}{mk\theta_M} \qquad [13]$$

Here, $h$ is the Planck constant, $T$ is the temperature of the measured region, $m$ is the mass of the silicon atom, $k$ is the Boltzmann constant, $\theta$ is the Bragg angle and λ is the x-ray wavelength. $\theta_M$ is the average Debye temperature given by $\frac{3}{\theta_M^2} = \frac{1}{\theta_l^2} + \frac{2}{\theta_t^2}$ where, $\theta_l$ and $\theta_t$ are Debye temperatures for the longitudinal and transverse modes of lattice vibrations. For silicon, $\theta_M = 543$ K [14]. $\left[\psi(x) + \frac{x}{4}\right]$ is the Debye function, where $x = \frac{\theta_M}{T}$. The value of the Debye function is close to unity for the temperature range, $T > \frac{\theta_M}{2}$ and is given by, $\left[\psi(x) + \frac{x}{4}\right] \approx 1 + \frac{x^2}{36} - \frac{x^4}{3600}$ [8].

At room temperature, the mean square atomic displacement, $\Delta X^2$, estimated by the Debye model is 0.0058 A°², and this compares extremely well with the $\Delta X^2$ calculated from the dispersion of Si using inelastic neutron scattering (0.0059 A°²) [15], and therefore, the Debye model can be reliably used to predict the temperature dependence of the structure factor.

For most materials, the integrated intensity of diffraction is directly proportional to the square of the structure factor, $|F|^2$. This is according to the kinematic theory, which states that the diffraction intensities from each small volume elements of a crystal add incoherently and the effects of multiple reflection of the same beam within the crystal need not be considered [12] [16]. Multiple reflections can be neglected for most crystals as imperfections resulting from lattice faults do not allow Bragg condition to be met for the reflected beam [12]. Kinematic diffraction results in a decrease in integrated intensity by the temperature factor, $e^{-2M}$, known as the Debye-Waller factor [8].

Silicon, however, can be free from defects or dislocations for large length scales [16]. Therefore high quality silicon can be considered a nearly perfect crystal, for which the reflected beams also satisfy the Bragg condition. The diffraction in silicon may require an explanation using the dynamical theory of diffraction, which takes into account the effect of multiple scattering. The integrated intensity in a dynamic diffraction is proportional to the amplitude of the structure factor, not the square of the structure factor as in the case of kinematic diffraction [8].

Batterman suggested two ways of interpreting the effect of thermal vibration in the case of dynamical diffraction [13]. The first is that multiple scattering still takes place in the case of thermal vibrations, but the decrease in scattering amplitude due to decrease in structure factor results in x-rays seeing more atoms, and in narrowing of the condition (width) for perfect refection (combined amplitude reflectivity approaching 1), also known as the Darwin width. The second way to interpret the effect of thermal vibration is that it degrades the perfection of the crystal, which would result in a kinematic diffraction from the crystal [13]. Using beams with bandwidth smaller than the expected Darwin widths, for (444), (555) and (660) reflections in thick and nearly perfect silicon crystal (impurities $\leq 10^{15}$ / cm$^3$), Batterman observed dynamical diffraction with peak narrowing that scaled linearly with the decrease in the structure factor, with no change in height of the rocking curve [14]. This resulted in a decrease in the intensity (integrated over the rocking curve) proportional to the decrease in structure factor, i.e., the integrated intensity decreased by the factor of $e^{-M}$. Kohra also observed an $e^{-M}$ dependence of the integrated intensity for dynamical diffraction from nearly perfect silicon and germanium crystals [17] [18].

In order to hypothesize if we expect dynamical diffraction in our experiment, one important parameter to look at is the x-ray extinction length. The extinction length is the path length through the crystal for which the amplitude reflectivity of the x-rays approach 1 and is given by $\Lambda_{ext} = \frac{1 v_c}{4 d r_0 |F|}$, where $d$ is the plane spacing, $r_0$ is the Thompson scattering length, and $v_c$ is the volume of the primitive unit cell [16]. For the current experimental conditions, for perfect silicon, the extinction lengths for (8 0 0), (12 0 0) and (16 0 0) are 15 μm (Sigma polarization), 51 μm (Sigma polarization) and 143 μm respectively [19]. These lengths are smaller than the x-ray path length in a single die in our experiment (192 μm, 128 μm and 365 μm respectively). Therefore, for a perfect silicon crystal, under the experimental conditions used (Section III), we expect dynamical diffraction.

The silicon used in the experiment, however, are grown by Czochralski (CZ) method, which have an oxide concentration of $10^{17}$- $10^{18}$ / cm$^3$ [20]; 2-3 orders of magnitude larger than the impurity concentration in nearly perfect silicon used by Kohra and Batterman. In transmission diffraction geometry, CZ grown silicon have shown significant deviation from dynamical theory, as dynamical diffraction is extremely susceptible to impurity induced strain [20]. For the high momentum transfers investigated in the experiment, impurity induced strains are expected to have an even more significant effect, as the per plane reflectivity is small and the extinction depths are large. Therefore, we could have results where diffraction significantly deviates from the dynamic theory and approaches the kinematic limit. The temperature dependence of the diffraction intensities should elucidate if the diffraction is kinematic or dynamic, as kinematic diffraction has a temperature dependence of $e^{-2M}$ while dynamic diffraction has a temperature dependence of $e^{-M}$.

## III. EXPERIMENTAL DETAILS

CZ grown undoped (1 0 0) silicon dies ($\rho > 400$ Ω.cm) with a thickness of 100 μm and area of 10 mm × 10 mm were obtained from UniversityWafer Inc. The stacking of two dies (Fig 1) was performed using a polyimide (PI) cage with a layer of 125 μm polyimide separating the two dies. The 125 μm polyimide used for the stacking was purchased from Sigma Aldrich (GF74134380) and was laser cut to obtain the cage and the separator.

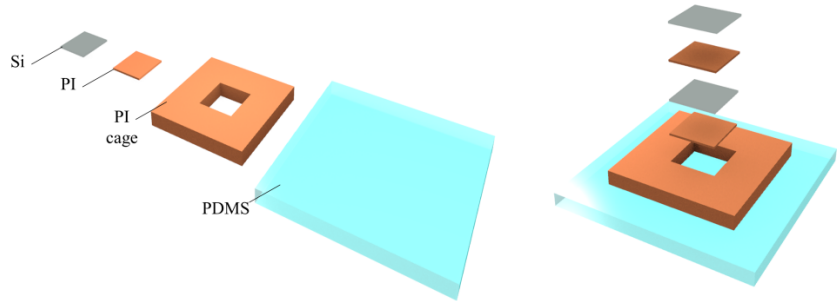

Fig 1. Schematic of the stacking of two silicon dies. The stacking was performed on top of a polydimethylsiloxane (PDMS) layer using a polyimide cage. A 125 μm polyimide layer separated the two silicon dies.

Temperature dependence of (8 0 0) and (12 0 0) diffractions (symmetric Bragg geometry) were measured on a single 100 μm silicon die on a Bruker D8 advance diffractometer at the School of Chemical Science at the University of Illinois. Temperature was controlled using Oxford Instruments M8 temperature controller. Mo Kα radiation (E = 17.45 keV) monochromatized with a Gobel type graded multilayer mirrors [21] to get an energy resolution of $\Delta E/E \approx 5 \times 10^{-2}$ and a beam divergence of ~5 mrad was incident on the sample with a flux of ~$10^7$ photons/s and a beam cross section of ~180 μm × 180 μm. Diffraction intensities were measured on a Bruker Photon II detector, and the diffraction rocking curves for (8 0 0) and (12 0 0) were recorded with a 4 deg sweep around the Bragg center with steps of 0.01 deg. An integration time of 4 s and 2 s were respectively used for each step. Detector files (.sfrm) were read using FabIO [22] library in python.

The measurement of the temperature dependence of a (16 0 0) diffraction and mapping of the diffraction (16 0 0) diffraction intensities were carried out at beamline 1-ID-E of the Advanced Photon Source, Argonne National Laboratory. Using a bent Si (1 1 1) double-Laue monochromator [23], the beam energy was selected as 72 keV, with a bandwidth ($\Delta E/E$) of 1.3 $\times 10^{-3}$ and a beam divergence of ~1.5 μrad. The wide bandwidth was chosen to relax the stacking accuracy required to meet the Bragg condition simultaneously for the two dies.

A flux of $5 \times 10^9$ photons/s was incident on an area of ~ 400 × 100 μm$^2$ on the sample, resulting in a spatial resolution of 400 × 100 μm$^2$. Diffraction intensities from (16 0 0) plane were collected by Pixirad-1 Cd-Te detector placed in symmetric Bragg geometry ($2\theta$ = 29.5 deg). The separation of the diffracted beams (~175 μm) being larger than Pixirad's pixel separation (60μm) allowed identification diffraction peaks from different dies.

Integrated intensity (sum of individual intensities at each point in a Bragg rocking curve) of the (16 0 0) reflection were collected from 121 points of the stacked die sample at room temperature. The sample was rotated by 0.05 degrees around the Bragg angle, with intensities recorded at each $5 \times 10^{-3}$ deg rotation with an integration time of 2 s per rotation. This resulted in the measurement time of one spot to be 20 s.

Temperature controlled measurements for the (16 0 0) were taken at 296 K, 321 K and 330 K from five different points of a single silicon die. The die was attached to the hot side of a Peltier cooler whose cold side was attached to a heat sink. The Peltier cooler was run at two different voltages to achieve equilibrium temperatures of 321 K and 330 K. A thermocouple attached on the surface of the Peltier module recorded the temperatures during the measurement.

## IV. RESULTS

Fig 2 shows the temperature dependence of integrated diffraction intensities for (8 0 0), (12 0 0) and (16 0 0) reflections respectively and Fig 3 compares the rocking curves at room temperature to those at 333 K (330 K for (16 0 0)). The temperature dependence for all the diffractions follows temperature dependence expected from kinematic theory.

The kinematic diffraction deduced from the temperature dependence is in contrast to the dynamic diffraction observed by Batterman and Kohra for nearly perfect silicon crystals. The result, however, is not surprising given CZ grown Si have significantly higher impurity concentration that those used by Batterman and Kohra.

Since the diffraction for all the investigated planes are kinematic, the sensitivity of diffraction intensity to temperature is double compared to the dynamic diffraction. For (16 0 0), the sensitivity is ~2.5, while for (12 0 0), the sensitivity is almost halved (~ 1.4) and is reduced by a factor of 4 (~0.65) for (8 0 0). Therefore, a sensitive thermometry using this method would require measuring symmetric Bragg reflections from a high momentum transfer plane similar to (16 0 0).

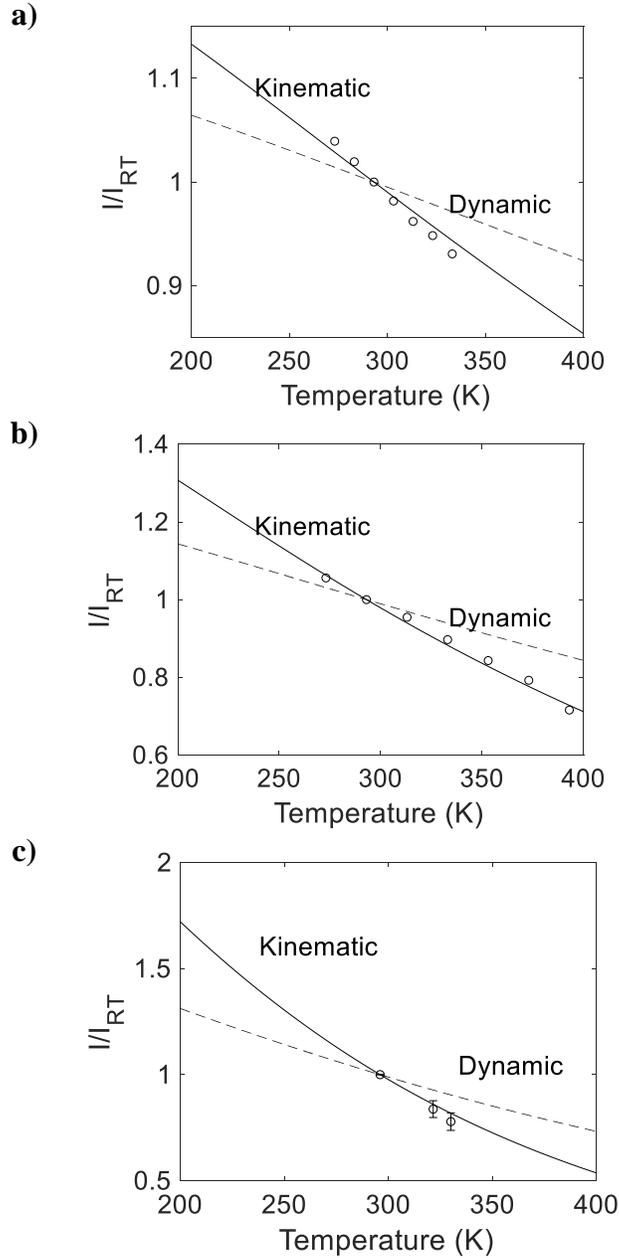

Fig 2. Temperature dependence of diffraction intensities for (a) (8 0 0), (b) (12 0 0) and (c) (16 0 0) planes. (8 0 0) and (12 0 0) are measured using a Bruker D8 instrument using Mo Kα radiation while (16 0 0) is measured using 72 keV synchrotron radiation at beamline 1-ID-E at the Advanced Photon Source. The diffraction intensities are normalized with the intensity at 293 K for (8 0 0) and (12 0 0) planes and at 296 K for (16 0 0) plane. The error bars for Fig 2(c) are calculated from measurements taken at 5 different spots in a die. The temperature dependence for each reflection is consistent with the expectation from the kinematic theory of diffraction.

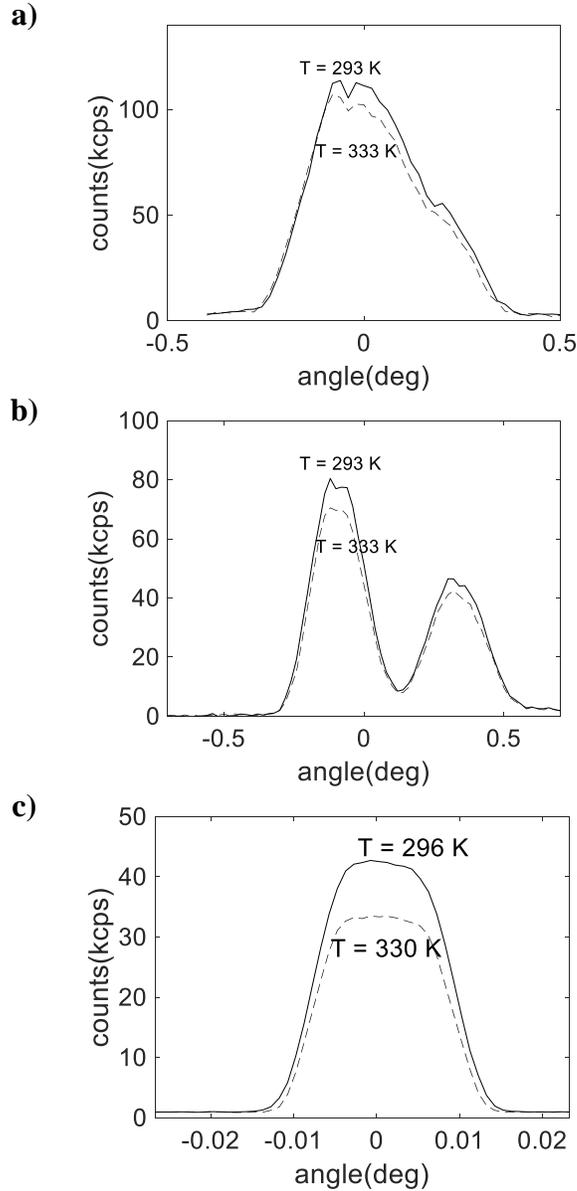

Fig 3. Rocking curves for (a) (8 0 0) and (12 0 0) planes at T = 293 K and T = 333 K and (c) (16 0 0) plane at T = 296 K and T = 330 K. Rocking curves (a) and (b) are obtained using Mo Kα beam while (c) uses 72 keV synchrotron source. The angles indicate rocking angles around the Bragg maximum. (12 0 0) shows two distinct Mo Kα$_1$ and Mo Kα$_2$ peaks while the two peaks are not well resolved for (8 0 0).  The decrease in width of the rocking curve is negligible compared to the decrease in the peak intensity.

An important demonstration using the synchrotron source was to show that the concurrent measurement and separation of diffraction intensities from different layers of stacked dies is possible, given the pixel size of the detector is smaller than the separation of the reflected beams. In the current experiment, the separation of the beam is given by:

$x = (2s \cot\theta - H \sin\theta)\cos\theta (\tan(2\theta) - \tan(\theta))$, where s is the distance between the two dies, and θ is the Bragg angle and $H \sin\theta$ is the spot size of a beam with a H x V cross section on the sample (the details of this calculations are included in the Supplemental Materials). For a Bragg angle of ~15 deg and the separation of dies of ~125 μm, the expected beam separation is ~175 μm along the x-axis of the detector (along the axis where the beam is elongated due to the small Bragg angle at incidence).

Fig 4 (b) shows the Pixirad image demonstrating that simultaneous observation of (16 0 0) reflections from two different dies was possible. The top and the bottom dies are identified due to the difference in their relative intensities due to x-ray absorption. The separation along the x-axis of the detector is similar to the expected value of ~ 175 μm. The wide bandwidth used in the experiment allowed observation of two different reflections in the same frame. Observation of reflections from different dies in the same frame is, however, not a general requirement, as a complete rocking curve should record diffraction intensities from each die even if they do not occur in the same frame.

a)

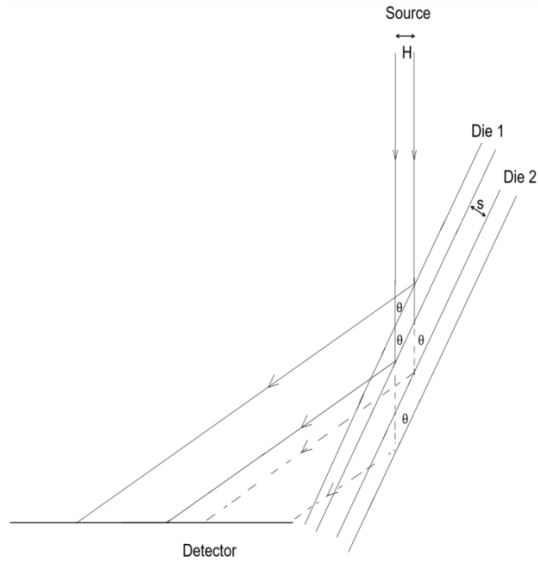

b)

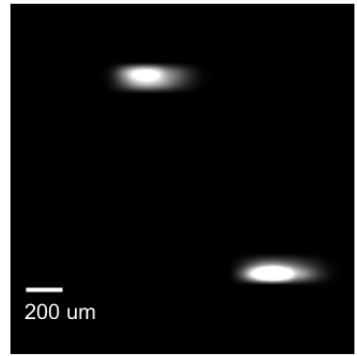

Fig 4. (a) Schematic of the experimental geometry used in beamline 1-ID-E. The x-axis separation of the beam from the top and bottom dies on the detector is a function of the physical separation of the two dies, 's' and the Bragg angle 'θ', and is given by
$x = (2s \cot\theta - H \sin\theta)\cos\theta (\tan(2\theta) - \tan(\theta))$. In the present experiment, the expected separation along x-axis is ~175 μm. (b) Pixirad image shows Bragg peaks from two dies simultaneously. The spot from the bottom die (top left hand corner) is less bright (about a factor of ~0.9) than that from the top die. This difference in intensity is expected due to absorption from the top silicon die (~7%) and the polyimide separator (~ 2%). The Pixirad image is consistent with the expected ~175 μm separation along x-axis . The separation along y-axis (~600 μm) is due to the imperfection in the azimuthal (χ) alignment of the dies (estimated to be ~0.6 mrad as the detector is ~1 m away from the sample) and depends on the sample to detector distance.

Fig 5 (a and b) show the of the distribution of integrated intensity at room temperature for the first and second dies in a stack. The rms deviation of the integrated intensity for different points of the sample was ~ 2.5% for the first die and 2.9% for the 2$^{nd}$ die. Since, the diffraction is kinematic, the sensitivity of diffraction intensity to temperature is ~2.5, and a 2.5% variation in intensity corresponds to a temperature uncertainty of ~ 3 K.

**a)** **b)**

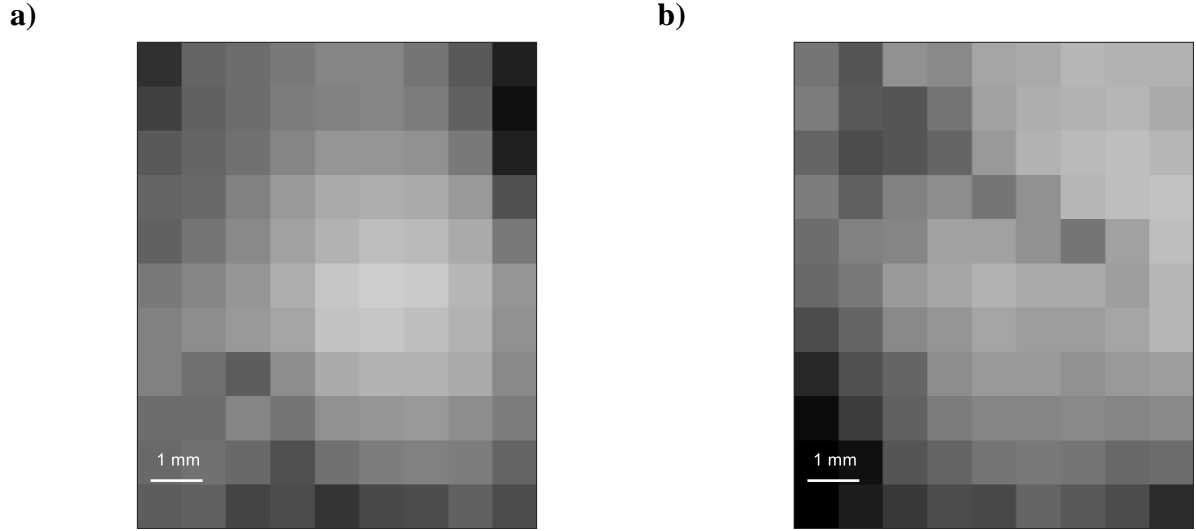

Fig 5. Distribution of the integrated intensity from 99 spots of (a) die 1 and (b) die 2 at room temperature. The colorscale represents the percentage deviation from the mean value of the integrated intensity, with black corresponging to -8% deviation and white corresponding to +8% deviation. The left and right ends of the die are not included in the map as the combined spot size and path length (~800 μm) is larger the distance from the edge of the die to the center of the beam (0.5 mm). The rms deviation in intensity at constant temperature is 2.5% and 3% respectively for die 1 and die 2. This contributes to the temperature uncertainty of 3 K and 3.5 K respectivey.

## V. DISCUSSION AND CONCLUSION

Our results show that simultaneous temperature mapping of different silicon dies in a stack is possible. We have demonstrated a mapping technique with a temperature resolution of 3 K spatial resolution of 400 μm × 100 μm (the spot size of the beam) and a time resolution of 20 s per spot.

We have observed that for high momentum transfer planes in an undoped silicon crystal grown by Czochralski (CZ) method, the diffraction in Bragg geometry, and consequently the temperature dependence of diffraction, is described well by the kinematic theory. Given the impurity (non Si) concentration in CZ silicon is similar for highly doped silicon (regardless of the method of growth), we expect kinematic theory applied in n- type dies irrespective of the method of their growth.

The spatial resolution of this method can be improved either by using a beam with a smaller cross section or by increasing the Bragg angle by using a beam of smaller energy, as long as the energy is sufficient to obtain a high momentum transfer peak ((16 0 0)) and have a absorption depth greater than the path length through the stack of dies.

The 2.5%-3% variation in intensity seems to be a systematic error rather than simply due to noise. Currently, we do not understand the cause of this variation. Obtaining a baseline image, which would be generally required for a 3D integrated circuit containing devices and wires, may reduce the temperature uncertainty coming from the intensity variation.

Finally, we note that our mapping experiment at Argonne (APS Beamline 1-ID-E) utilizes several desirable properties of a synchrotron source: an x-ray beam with high energy and intensity and a small beam spot size, and these properties are required for a sensitive and accurate temperature mapping. High energy x-rays are required to obtain diffraction intensities for high momentum transfer peak like (16 0 0). High momentum transfer is desirable for sensitive thermometry as the Debye-Waller factor scales as $\left(\frac{1}{d}\right)^2$ [8]. An energy of 72 keV, however, is not a requirement to obtain a diffraction peak for (16 0 0). In fact, any x-ray energy above ~19 keV can satisfy the Bragg condition to obtain a (16 0 0) peak. High photon flux is required for obtaining high enough diffraction intensity for accurate temperature mapping. Therefore, compact Compton sources (CCS) [24] or x-ray tubes with liquid metal anodes [25] that provide about two orders of magnitude higher flux than conventional x-ray tubes (~$10^8$ photons/s) and a monochromatic beam of energy >20 keV are required to use this technique in a laboratory setting.

## ACKNOWLEDGEMENTS

The authors thank Dr. Toby Woods of the School of Chemical Sciences, University of Illinois for assistance with the experimental setup and data collection for the diffraction experiments. The authors also thank Prof. Seok Kim and Dr. Jun Kyu Park of the Department of Mechanical Science and Engineering, University of Illinois for help with the stacking of silicon dies. This research was funded by Semiconductor Research Corporation (Task ID: 3044.0001). This research used resources of the Advanced Photon Source, a U.S. Department of Energy (DOE) Office of Science User Facility operated for the DOE Office of Science by Argonne National Laboratory under Contract No. DE-AC02-06CH11357.